**Density functional theory analysis of the interplay between Jahn-Teller instability, uniaxial magnetism, spin arrangement, metal-metal interaction and spin-orbit coupling in $Ca_3CoMO_6$ (M = Co, Rh, Ir)**

Yuemei Zhang[1], Erjun Kan[1], Hongjun Xiang[2], Antoine Villesuzanne[3] and Myung-Hwan Whangbo[1,*]

[1] Department of Chemistry, North Carolina State University, Raleigh, NC 27695-8204, USA

[2] Key Laboratory of Computational Physical Sciences (Ministry of Education) and Department of Physics, Fudan University, Shanghai 200433, P. R. China

[3] CNRS, Université de Bordeaux, ICMCB, 87 Av. Dr. A. Schweitzer, 33608 Pessac cedex, France

**Abstract**

In the isostructural oxides $Ca_3CoMO_6$ (M = Co, Rh, Ir), the $CoMO_6$ chains made up of face-sharing $CoO_6$ trigonal prisms and $MO_6$ octahedra are separated by Ca atoms. We analyzed the magnetic and electronic properties of these oxides on the basis of density functional theory calculations including on-site repulsion and spin-orbit coupling, and examined the essential one-electron pictures hidden behind results of these calculations. Our analysis reveals an intimate interplay between Jahn-Teller instability, uniaxial magnetism, spin arrangement, metal-metal interaction, and spin-orbit coupling in governing the magnetic and electronic properties of these oxides. These oxides undergo a Jahn-Teller distortion but their distortions are weak, so that their trigonal-prism $Co^{n+}$ (n = 2, 3) ions still give rise to strong easy-axis anisotropy along the chain direction. As for the d-state split pattern of these ions, the electronic and magnetic properties of $Ca_3CoMO_6$ (M = Co, Rh, Ir) are consistent with $d_0 < (d_2, d_{-2}) < (d_1, d_{-1})$, but not with $(d_2, d_{-2}) < d_0 < (d_1, d_{-1})$. The trigonal-prism $Co^{3+}$ ion in $Ca_3Co_2O_6$ has the L = 2 configuration $(d_0)^1(d_2, d_{-2})^3(d_1, d_{-1})^2$ because of the metal-metal interaction between adjacent $Co^{3+}$ ions in each $Co_2O_6$ chain, which is mediated by their $z^2$ orbitals, and the spin-orbit coupling of the trigonal-prism $Co^{3+}$ ion. The spins in each $CoMO_6$ chain of $Ca_3CoMO_6$ prefer the ferromagnetic arrangement for M = Co and Rh, but the antiferromagnetic arrangement for M = Ir. The octahedral $M^{4+}$ ion of $Ca_3CoMO_6$ has the $(1a)^1(1e)^4$ configuration for M = Rh but the $(1a)^2(1e)^3$ configuration for M = Ir, which arises from the difference in the spin-orbit coupling of the $M^{4+}$ ions and the Co…M metal-metal interactions.

**1. Introduction**

For a magnetic system with transition-metal ions exhibiting uniaxial (i.e., Ising) magnetism, the ions have an unevenly filled degenerate d-states so that the $\Delta J_z$ value of the lowest-lying Kramer's doublet state is greater than 1.[1] However, such a system has Jahn-Teller (JT) instability [2] and the associated JT distortion may lift the d-state degeneracy causing the uniaxial magnetism. Thus, a true uniaxial magnetism is not possible unless a JT distortion is prevented by steric hindrance.[3] Indeed, uniaxial magnetism and JT instability were found to compete in the magnetic oxide $Ca_3CoMnO_6$,[4] which consists of the $CoMnO_6$ chains made up of face-sharing $CoO_6$ trigonal prism (TP) and $MnO_6$ octahedron (OCT) units (**Fig. 1**). In the room-temperature structure of $Ca_3CoMnO_6$,[5] each $CoMnO_6$ chain has a three-fold rotational symmetry, $C_3$, so that the high-spin $Co^{2+}$ ($d^7$) ion at each TP $CoO_6$ has the d-electron configuration $(d_0)^2(d_2, d_{-2})^3(d_1, d_{-1})^2$, giving rise to both JT instability and uniaxial magnetism. Here we use the local coordinate system in which the z-axis is taken along the $CoMO_6$ chain (i.e., the crystallographic c-direction), so that the $d_0$ orbital is equivalent to the $z^2$ orbital, the degenerate $(d_2, d_{-2})$ set to the $(x^2-y^2, xy)$ set, and the degenerate $(d_1, d_{-1})$ set to the $(xz, yz)$ set (**Fig. 2a**).[1] $Ca_3CoMnO_6$ is regarded to have uniaxial spins,[6] but first principles density functional theory (DFT) calculations showed [4] that $Ca_3CoMnO_6$ should undergo a JT distortion removing the $C_3$ symmetry and hence cannot be truly uniaxial, although it has strong magnetic anisotropy with the easy axis along the $CoMnO_6$ chain.

The magnetic oxides $Ca_3CoMO_6$ (M = Co,[7] Rh,[8] Ir [9]), isostructural with $Ca_3CoMnO_6$,[5] belongs to the family of hexagonal perovskites.[10] Due to the face sharing of the TP $CoO_6$ and OCT $MO_6$, the nearest-neighbor (NN) Co…M distance of the $CoMO_6$ chain is short (i.e., 2.595, 2.682 and 2.706 Å for M = Co, Rh and Ir, respectively) so that the Co…M direct metal-metal interaction mediated by their $z^2$ orbitals can be substantial. (Here the NN Co…Ir distance of 2.706 Å is taken from the structure of $Ca_3CoIrO_6$ optimized by DFT calculations. See below.) The X-ray photoemission study [11] of $Ca_3CoMO_6$ revealed that the Co atoms of the TP $CoO_6$ exist as $Co^{3+}$ ions for M = Co but as $Co^{2+}$ ions for M = Rh and Ir, and hence the M atoms of the $MO_6$ octahedra exist as $M^{3+}$ ions for M = Co but as $M^{4+}$ ions for M = Rh and Ir. The magnetic properties of $Ca_3CoMO_6$ (M = Co, Rh, Ir) show that the TP Co atoms are present as high-spin ions, the $CoMO_6$ chains have uniaxial spins, and their intrachain spin arrangement is ferromagnetic (FM) for M = Co [12-14] and Rh, [15] and the same is presumed to be true for M = Ir.[16] The electronic and magnetic properties of $Ca_3CoMO_6$ (M = Co, Rh, Ir) have been investigated in a number of DFT studies.[17-25]

It has been well established that the d-states of a transition metal ion at an isolated TP site with $C_3$ rotational symmetry are split as $d_0 < (d_2, d_{-2}) < (d_1, d_{-1})$.[26] This leads to L = 0 configuration $(d_0)^2(d_2, d_{-2})^2(d_1, d_{-1})^2$ (**Fig. 3a**) for an isolated TP high-spin $Co^{3+}$ ($d^6$) ion, hence predicting the absence of uniaxial magnetism. Thus, it was concluded [1] that the TP $Co^{3+}$ ion of $Ca_3Co_2O_6$ should have the L = 2 configuration $(d_0)^1(d_2, d_{-2})^3(d_1, d_{-1})^2$ (**Fig. 3b**) due to the interaction between the $z^2$ orbitals of adjacent TP and OCT $Co^{3+}$ ions. In the DFT study of $Ca_3Co_2O_6$ by Wu et al.,[21] spin-orbit coupling (SOC) interactions were found essential for the occurrence of the L = 2 configuration $(d_0)^1(d_2, d_{-2})^3(d_1, d_{-1})^2$; the TP $Co^{3+}$ ion has the $(d_0)^1(d_2, d_{-2})^3(d_1, d_{-1})^2$ configuration if SOC interactions are included, but the $(d_0)^2(d_2, d_{-2})^2(d_1, d_{-1})^2$ configuration otherwise. Nevertheless, they assumed the split pattern of the TP $Co^{3+}$ ion to be $(d_2, d_{-2}) < d_0 < (d_1, d_{-1})$, which leads to the L = 2 configuration $(d_2, d_{-2})^3(d_0)^1(d_1, d_{-1})^2$ (**Fig. 3c**) even if the SOC effect is not included. Furthermore, Burnus et al.[27] employed this L = 2 configuration for the TP $Co^{3+}$ ion to interpret their X-ray absorption and X-ray magnetic dichroism data of $Ca_3Co_2O_6$, and concluded that the $d_0 < (d_2, d_{-2}) < (d_1, d_{-1})$ pattern is incorrect for the TP $Co^{3+}$ ion. However, the $(d_2, d_{-2}) < d_0 < (d_1, d_{-1})$ split pattern gives rise to serious conceptual difficulties. First, for a transition metal atom surrounded by oxygen ligands, the split pattern of its the d-states is determined by how strong the antibonding interactions between the metal $n$d and O 2p orbitals are.[28] The $z^2$ orbital of the TP $Co^{3+}$ ion, being aligned along the $C_3$ axis of the TP $CoO_6$, overlaps least well with the 2p-orbitals of the surrounding O atoms. As a consequence, the $d_0$ level should be the lowest-lying state of the TP Co d-states (**Fig. 2a**) regardless of whether the TP ion is $Co^{3+}$ or $Co^{2+}$, so that the $(d_2, d_{-2}) < d_0 < (d_1, d_{-1})$ split pattern cannot be correct. Second, the $(d_2, d_{-2}) < d_0 < (d_1, d_{-1})$ split pattern cannot explain the uniaxial magnetism of $Ca_3CoRhO_6$, because it gives rise to the L = 0 configuration $(d_2, d_{-2})^4(d_0)^1(d_1, d_{-1})^2$ for the TP $Co^{2+}$ ($d^7$) ion (**Fig. 3d**). In contrast, the $d_0 < (d_2, d_{-2}) < (d_1, d_{-1})$ split pattern gives the L = 2 configuration $(d_0)^2(d_2, d_{-2})^3(d_1, d_{-1})^2$ (**Fig. 3e**), and the latter is consistent with the density functional calculations for $Ca_3CoRhO_6$ by Wu et al.[25] In interpreting their X-ray absorption and X-ray magnetic dichroism data of $Ca_3CoRhO_6$, Burnus et al.[29] used the $d_0 < (d_2, d_{-2}) < (d_1, d_{-1})$ pattern for the TP $Co^{2+}$ ion and suggested that the $d_0$ state is nearly degenerate with the $(d_2, d_{-2})$ states.

The above discussion raises several important questions: (a) It is necessary to determine whether or not the spins of the $CoIrO_6$ chains in $Ca_3CoIrO_6$ have the FM arrangement as found in $Ca_3Co_2O_6$ and $Ca_3CoRhO_6$. (b) In all three oxides $Ca_3CoMO_6$ (M = Co, Rh, Ir), the TP $Co^{n+}$ (n = 2 or 3) ions possess the L = 2 electron configuration. Thus, $Ca_3CoMO_6$ (M = Co, Rh, Ir) should be susceptible to JT instability as found for $Ca_3CoMnO_6$.[4] It is important to examine how strong their JT distortions can be. (c) Concerning the d-state split pattern of a transition-metal ion at a TP site, it is controversial whether the $d_0 < (d_2, d_{-2}) < (d_1, d_{-1})$ or $(d_2, d_{-2}) < d_0 < (d_1, d_{-1})$ pattern is correct. It is desirable to determine if the split pattern depends on the charge of the TP $Co^{n+}$ (n = 2, 3) ion as reported in the studies of Burnus et al.[27,29] and/or whether the $d_0$ state is nearly degenerate with the $(d_2, d_{-2})$ states as suggested by Burnus et al.[29] (d) In $Ca_3CoMO_6$ (M = Rh, Ir), the OCT $M^{4+}$ ions might exhibit SOC effects because of their open-shell electron configuration $(t_{2g})^5$. It is interesting to examine if the SOC effects of these ions affect the electronic and magnetic structures of $Ca_3CoMO_6$. In the present work, we investigate these questions on the basis of DFT calculations for $Ca_3CoMnO_6$ (M = Co, Rh, Ir). Results of our study are presented in what follows.

## 2. Calculations

To optimize the crystal structures of $Ca_3CoMO_6$ (M = Co, Rh, Ir) in the presence and absence of $C_3$ rotational symmetry, we employed the projector augmented wave (PAW) method encoded in the Vienna ab initio simulation package (VASP)[30] with the local spin density approximation (LSDA). To properly describe the electron correlation associated with the d states of transition metal atoms, the LSDA plus on-site repulsion U (LSDA+U) method was adopted.[31] In addition, SOC effects[32] were considered by performing LSDA+U+SOC calculations with the spins oriented parallel and perpendicular to the $CoMO_6$ chain direction (hereafter the //c- and ⊥c-spin orientations, respectively). The convergence threshold for our LSDA+U+SOC calculations was set to $10^{-5}$ eV in energy and $10^{-2}$ eV/Å in force with the plane-wave cutoff energy of 400 eV and a set of 3×3×3 k-points for the irreducible Brillouin zone. For $Ca_3CoIrO_6$, only the cell parameters have been reported.[33] Therefore, we determined the atomic positions of $Ca_3CoIrO_6$ by optimizing the crystal structure on the basis of LSDA+U+SOC calculations. This optimization leads to two kinds of structures for each $Ca_3CoMO_6$ (M = Co, Rh, Ir), namely, one with high orbital moment ($\mu_L$), and the other with low $\mu_L$, on the TP $Co^{n+}$ (n = 2, 3) ions. As found for $Ca_3CoMnO_6$,[4] the $CoMO_6$ chains of $Ca_3CoMO_6$ have the $C_3$-rotational symmetry in the high-$\mu_L$ structure, but do not in the low-$\mu_L$ structure. The geometry optimization with LSDA+U+SOC calculations is carried out with no symmetry constraint, so it is generally difficult to have the calculations converge to the high-$\mu_L$ structure.

In discussing the spin and orbital moments of the TP and OCT ions of $Ca_3CoMO_6$ (M = Co, Rh, Ir) as well as their density of states (DOS), we have carried out LSDA+U+SOC calculations for the experimental and the optimized structures of $Ca_3Co_2O_6$ and $Ca_3CoRhO_6$ and for the optimized structure of $Ca_3CoIrO_6$ by using the full-potential linearized augmented plane wave (FPLAPW) method [34] encoded in the WIEN2k package,[35] 5×5×5 k-points for the irreducible Brillouin zone, the threshold of $10^{-5}$ Ry for the energy convergence, the cut-off energy parameters of $RK_{max}$ = 7 and $G_{max}$ = 12, and the energy threshold of −9.0 Ry for the separation of the core and valence states.

For the effective on-site repulsion $U_{eff} = U - J$ (where J is the Stoner intra-atomic parameter) needed for the geometry optimization with the LSDA+U+SOC (VASP) calculations, we used $U_{eff} = 4$ eV on Co for $Ca_3Co_2O_6$, $U_{eff} = 4$ eV on Co and $U_{eff} = 2$ eV for Rh and Ir for $Ca_3CoMO_6$ (M = Rh, Ir). (We note that LSDA+U and LSDA+U+SOC calculations, only the difference $U - J = U_{eff}$ matters for the calculations.) These parameters are quite similar to those employed by Wu *et al.* in their DFT studies of $Ca_3Co_2O_6$ [21] and $Ca_3CoRhO_6$.[25] We also employed these parameters for our LSDA+U+SOC (WIEN2k) calculations on $Ca_3CoMO_6$ (M = Co, Rh, Ir) to find that the use of $U_{eff} = 4$ and 2 eV on Co and M, respectively, does not lead to magnetic insulating states for $Ca_3CoMO_6$ (M = Rh, Ir), but the use of $U_{eff} = 4$ eV on both Co and M does. Hereafter, the $U_{eff}$ values on Co, Rh and Ir will be designated as $U_{eff}(Co)$, $U_{eff}(Rh)$ and $U_{eff}(Ir)$, respectively.

Possible ordered spin arrangements for each $CoMO_6$ chain of $Ca_3CoMO_6$ (M = Co, Rh, Ir) include the FM (i.e., ↑↑↑↑), antiferromagnetic (AFM) (i.e., ↑↓↑↓) and ↑↑↓↓ arrangements. It should be noted that the AFM state represents a ferrimagnetic arrangement in each $CoMO_6$ chain because the magnetic moments of the Co and M sites are different (see below). In our calculations, the spin arrangement between adjacent $CoMO_6$ chains is assumed to be FM.

## 3. Magnetic ground states of $Ca_3CoM_6$ (M = Co, Rh, Ir)

Our WIEN2k calculations show that, for $Ca_3Co_2O_6$, a magnetic insulating state can be obtained at the LSDA+U and LSDA+U+SOC levels of calculations, but the LSDA+U+SOC level of calculations are necessary to obtain the L = 2 configuration $(d_0)^1(d_2, d_{-2})^3(d_1, d_{-1})^2$ for the TP $Co^{3+}$ ion, as found by Wu et al.[21] For both $Ca_3CoRhO_6$ and $Ca_3CoIrO_6$, a magnetic insulating state is obtained only at the LSDA+U+SOC level of calculations. Our LSDA+U+SOC calculations reveal that only the FM state is stable for the experimental structure of $Ca_3Co_2O_6$, while both the FM and ↑↑↓↓ states are stable for the experimental structure of $Ca_3CoRhO_6$. The FM state is more stable than the ↑↑↓↓ state by 308 meV per formula unit (FU) from calculations with $U_{eff}(Co) = 4$ eV and $U_{eff}(Rh) = 2$ eV, and by 422 meV/FU from calculations with $U_{eff}(Co) = U_{eff}(Rh) = 4$ eV. For the optimized structure of $Ca_3CoIrO_6$ with $C_3$ symmetry, only the AFM state is stable as long as $U_{eff}(Co) \geq U_{eff}(Ir)$ in LSDA+U+SOC calculations. When $U_{eff}(Co) < U_{eff}(Ir)$, LSDA+U+SOC calculations lead to a stable FM state, but the FM state is less stable than the AFM state (e.g., by 39 meV/FU with $U_{eff}(Co) = 3.5$ eV and $U_{eff}(Ir) = 4.0$ eV).

**4. Jahn-Teller distortion and magnetic anisotropy**

To see whether $Ca_3CoMO_6$ (M = Co, Rh, and Ir) undergoes a JT distortion, all the structures of $Ca_3CoMO_6$ (M = Co, Rh, and Ir) were optimized by performing LSDA+U+SOC (VASP) calculations with the //c-spin orientation for their FM states. In the geometry optimizations, the cell parameters were fixed at the experimental values, but the atom positions were allowed to relax with and without the $C_3$ rotational symmetry for each $CoMO_6$ chain. The atom positions of the optimized structures of $Ca_3CoMO_6$ (M = Co, Rh, Ir) are summarized in **Tables S1 – S3** of the supporting information.

The relative energies of the experimental and optimized structures of $Ca_3CoMO_6$ (M = Co, Rh, Ir) obtained by LSDA+U+SOC (VASP) calculations are summarized in **Table 1**, and the spin and orbital moments ($\mu_S$ and $\mu_L$, respectively) of the TP and OCT transition-metal ions of $Ca_3CoMO_6$ (M = Co, Rh, Ir) obtained by the LSDA+U+SOC (WIEN2k) calculations in **Tables 2 – 4**. For the optimized structure with $C_3$ rotational symmetry, the orbital moment on the TP $Co^{n+}$ ion is large (i.e., $\mu_L$ = 1.48, 1.68 and 1.69 $\mu_B$ for M = Co, Rh and Ir, respectively). However, for the optimized structure without $C_3$ symmetry, the orbital moment on the TP $Co^{n+}$ ion is smaller (i.e., $\mu_L$ = 0.47, 0.60 and 0.30 $\mu_B$ for M = Co, Rh and Ir, respectively). (The spin and orbital moments of the OCT sites will be discussed in Section 6.) For each $Ca_3CoMO_6$ (M = Co, Rh or Ir), the optimized structure without $C_3$ symmetry is more stable than that with $C_3$ symmetry (i.e., $\Delta E$ = 65.5, 35.2 and 139.8 meV per FU for M = Co, Rh and Ir, respectively). This shows that the structural change of $Ca_3CoMO_6$ (M = Co, Rh or Ir), from the structure with $C_3$ symmetry to that without $C_3$ symmetry, is a JT distortion. **Fig. 4** shows the atom displacements involved in the JT distortions of $Ca_3CoMO_6$ (M = Co, Rh, Ir), with respect to the experimental structure for M = Co and Rh, and with respect to the optimized structure with $C_3$-rotational symmetry for M = Ir. In $Ca_3Co_2O_6$ with TP $Co^{3+}$ ions, the largest displacement (0.064 Å) is found for one of the O atoms with a smaller displacement for the TP Co atom (i.e., 0.027 Å). In $Ca_3CoRhO_6$ and $Ca_3CoIrO_6$ with TP $Co^{2+}$ ions, however, the TP Co atom shows the largest displacement (i.e., 0.064 and 0.051 Å, respectively). A probable cause for this difference is discussed in Section 7.

An important consequence of the JT distortion is that the orbital moments of the TP $Co^{n+}$ ions are reduced by the JT distortion but the JT distortions are not strong enough to completely quench the orbital angular moment of $Co^{n+}$ (**Tables 2 – 4**). As found for $Ca_3CoMnO_6$,[4] therefore, the oxides $Ca_3CoMO_6$ (M = Co, Rh, Ir) cannot possess a genuine uniaxial magnetism. We investigate the preference of their spin orientation by performing LSDA+U+SOC (WIEN2k) calculations for the JT-distorted $Ca_3Co_2O_6$, $Ca_3CoRhO_6$ and $Ca_3CoIrO_6$ with //c- and ⊥c-spin orientations. Our calculations show that the ⊥c-spin orientation is less stable than the //c-spin orientation by 33, 26 and 27 meV/FU for $Ca_3Co_2O_6$, $Ca_3CoRhO_6$ and $Ca_3CoIrO_6$, respectively, which represent very strong easy-axis anisotropy. This renders the observed anisotropic magnetic character to $Ca_3Co_2O_6$, $Ca_3CoRhO_6$ and $Ca_3CoIrO_6$.

## 5. One-electron picture in DFT+U description: Analysis of the electronic structure of $Ca_3Co_2O_6$

In general, it is not straightforward to decipher a one-electron picture hidden behind the results of DFT calculations especially when the latter include effects of spin-polarization/on-site repulsion.[36] To estimate the energy separation $\Delta_1$ between $d_0$ and ($d_2$, $d_{-2}$) as well as the energy separation $\Delta_2$ between ($d_2$, $d_{-2}$) and ($d_1$, $d_{-1}$) (see **Fig. 2a**) of the TP $Co^{3+}$ ion in $Ca_3Co_2O_6$, one may perform one-electron tight binding-calculations for an isolated TP $CoO_6$. The $CoO_6$ TP found in $Ca_3Co_2O_6$ differs slightly from the ideal $CoO_6$ TP in that the two $O_3$ triangular faces are not eclipsed but are rotated away from each other around the $C_3$ axis by the angle $\phi = 14.25°$. Our extended Hückel tight-binding calculations[37,38] for the $CoO_6$ TP show that $\Delta_1 = 0.20$ eV and $\Delta_2 = 0.65$ eV for the $CoO_6$ TP with $\phi = 0°$, while $\Delta_1 = 0.13$ eV and $\Delta_2 = 0.74$ eV for the $CoO_6$ TP with $\phi = 14.25°$. (The atomic parameters used for these calculations are summarized in **Table S4** of the supporting information.) Thus, $\Delta_1$ (0.13 – 0.20 eV) is greater than the typical SOC energy expected for 3d transition metal oxides (i.e., less than 0.05 eV)[39] by a factor of 3 – 4, and is smaller than $\Delta_2$ only by a factor of 3 – 6. Consequently, the correct d-state split pattern for the TP Co$^{n+}$ ions of Ca$_3$CoMO$_6$ should be $d_0 < (d_2, d_{-2}) < (d_1, d_{-1})$ with $\Delta_1 = 0.13 - 0.20$ eV and $\Delta_2 = 0.65 - 0.74$ eV. In the following we examine how this split pattern is manifested in the LSDA+U and LSDA+U+SOC calculations for Ca$_3$Co$_2$O$_6$.

As reported by Wu et al.,[21] our study for Ca$_3$Co$_2$O$_6$ shows that the TP Co$^{3+}$ ion has the L = 2 configuration $(d_0)^1(d_2, d_{-2})^3(d_1, d_{-1})^2$ in the LSDA+U+SOC calculations but the L = 0 configuration $(d_0)^2(d_2, d_{-2})^2(d_1, d_{-1})^2$ in the LSDA+U calculations. This can be seen from the projected DOS plots presented in **Fig. 5**. To understand the switching of the L = 0 configuration to the L = 2 configuration by the action of SOC, it is necessary to consider three effects, i.e., the spin arrangement between adjacent TP and OCT Co$^{3+}$ ions,[1,40] the direct metal-metal interaction between them, and the SOC on the TP Co$^{3+}$ ion.[21,40] It is convenient to discuss these factors by considering an isolated dimer made up of adjacent TP CoO$_6$ and OCT CoO$_6$, as pointed out elsewhere.[40]

We first consider the interaction between the $z^2$ orbitals of adjacent Co$^{3+}$ ions. In a one-electron tight-binding description, the high-spin Co$^{3+}$ (d$^6$) ion of an isolated TP CoO$_6$ has the $(d_0)^2(d_2, d_{-2})^2(d_1, d_{-1})^2$ configuration while the low-spin Co$^{3+}$ (d$^6$) ion of an isolated OCT CoO$_6$ has the $(t_{2g})^6$ configuration. The OCT CoO$_6$ in Ca$_3$Co$_2$O$_6$ has C$_3$ symmetry, so the t$_{2g}$ level is split into the 1a and 1e set as depicted in **Fig. 2b**. The $z^2$ orbital of the TP Co$^{3+}$ ion can interact directly with the 1a orbital (i.e., the $z^2$ orbital) of the OCT Co$^{3+}$ ion through the shared triangular face due to the very short NN Co…Co distance (2.595 Å). In describing such an interaction at the spin-polarized DFT+U level, it should be noted that one-electron energy levels given by tight-binding calculations are split into the up-spin and down-spin levels by the spin-polarization/on-site repulsion, as illustrated in **Fig. 6**. Thus, the L = 0 configuration $(d_0)^2(d_2, d_{-2})^2(d_1, d_{-1})^2$ of the TP $Co^{3+}$ ion means that the LUMO of the TP $CoO_6$ is given by the $(d_2, d_{-2})\downarrow$ level. Therefore, if one of the four electrons present in the two $z^2$ orbitals of adjacent TP and OCT $Co^{3+}$ ions is transferred to the $(d_2, d_{-2})\downarrow$ level of the TP $Co^{3+}$ ion, the resulting electron configuration of the TP $Co^{3+}$ ion would be close to $(d_0)^1(d_2, d_{-2})^3(d_1, d_{-1})^2$. The spins of the TP and OCT $Co^{3+}$ ions are assumed to have the FM arrangement in **Fig. 7a**, where the $z^2\uparrow$ and $z^2\downarrow$ levels of the OCT $Co^{3+}$ ion are split less than those of the TP $Co^{3+}$ ion because, to a first approximation, the OCT site has a low-spin $Co^{3+}$ ion whereas the TP site has a high-spin $Co^{3+}$ ion. Since both TP and OCT sites have $Co^{3+}$ ions, the midpoint between their $z^2\uparrow$ and $z^2\downarrow$ levels should be nearly the same. The highest occupied level resulting from the $z^2$ orbitals of the two $Co^{3+}$ ions is the $\sigma*\downarrow$ level, in which the weight of the TP $z^2\downarrow$ orbital is larger than the OCT $z^2\downarrow$ orbital because the former lies higher in energy than the latter. In the DFT+U level of description, the occupied $\sigma*\downarrow$ level lies below the empty $(d_2, d_{-2})\downarrow$ level of the TP $Co^{3+}$ ion. The effect of the SOC interaction at the TP $Co^{3+}$ ion site is depicted in **Fig. 8a**, where the SOC splits the unoccupied degenerate level $(d_2, d_{-2})\downarrow$ into the $d_2\downarrow$-below-$d_{-2}\downarrow$ pattern since $\lambda < 0$ for $Co^{3+}$ ($d^6$). When the unoccupied $d_2\downarrow$ level is lowered below the occupied $\sigma*\downarrow$ level, an electron transfer occurs from the $\sigma*\downarrow$ level to the $d_2\downarrow$ level. Since the $\sigma*\downarrow$ level has a greater weight on the TP $z^2\downarrow$ orbital, this charge transfer effectively amounts to the configuration switch of the TP $Co^{3+}$ from the L = 0 configuration $(d_0)^2(d_2, d_{-2})^2(d_1, d_{-1})^2$ to the L = 2 configuration $(d_0)^1(d_2, d_{-2})^3(d_1, d_{-1})^2$. This is why the TP $Co^{3+}$ ion has the $(d_0)^2(d_2, d_{-2})^2(d_1,$ $d_{-1})^2$ configuration at the DFT+U level, but has the $(d_0)^1(d_2, d_{-2})^3(d_1, d_{-1})^2$ configuration at the DFT+U+SOC level. This explanation is based on the d-state split pattern of $d_0 < (d_2, d_{-2}) < (d_1, d_{-1})$ for the TP $Co^{3+}$ ion. If the TP $Co^{3+}$ ion were to have the $(d_2, d_{-2}) < d_0 < (d_1, d_{-1})$ split pattern (**Fig. 8b**), the TP $Co^{3+}$ ion would have the $(d_2, d_{-2})^3(d_0)^1(d_1, d_{-1})^2$ configuration in both DFT+U and DFT+U+SOC levels of descriptions because the $\sigma^*\downarrow$ level remains unoccupied regardless of whether or not the singly occupied $(d_2, d_{-2})\downarrow$ level is split by the effect of SOC.

## 6. Electronic structures of $Ca_3CoMO_6$ (M = Rh, Ir)

The reason why $Ca_3CoMO_6$ (M = Rh, Ir) has $Co^{2+}$ and $M^{4+}$ ions in the TP and OCT sites is that the Co 3d orbital lies lower in energy, and is more contracted, than the Rh 4d and Ir 5d orbitals. Thus, the essential features of the direct metal-metal interaction between the TP $Co^{2+}$ and OCT $Rh^{4+}$ ions, which give rise to the configuration $(z^2)^2(x^2-y^2, xy)^3(xz, yz)^2$ for the TP $Co^{2+}$ ion and the configuration $(1a)^1(1e)^4$ for the OCT $Rh^{4+}$ ion, can be understood in terms of the orbital interaction diagram shown in **Fig. 9a**. Here the two adjacent ions have the FM arrangement, and the midpoint between the $z^2\uparrow$ and $z^2\downarrow$ orbitals is placed higher in energy for the $M^{4+}$ (M = Rh, Ir) ion than that for the $Co^{2+}$ ion, because the Rh 4d and Ir 5d orbital is more diffuse, and lies higher in energy, than the Co 3d orbital.[38,41] The $\sigma^*\downarrow$ level (i.e., the highest-lying level arising from the interactions between the $z^2$ orbitals of adjacent $Co^{2+}$ and $M^{4+}$ site) has a larger weight on the OCT $M^{4+}$ ion so that the absence of an electron in the $\sigma^*\downarrow$ level amounts to the $(1a)^1(1e)^4$ configuration for the OCT $M^{4+}$ ion.

The low-spin OCT $M^{4+}$ ($d^5$) ion of $Ca_3CoMO_6$ (M = Rh, Ir) has the open-shell configuration, $(t_{2g})^5$, and the Rh and Ir atoms are a heavier element than Co. Therefore, the local electronic structure of $M^{4+}$ ($d^5$) ion can be more strongly affected by the SOC compared with that of the OCT $Co^{3+}$ ($d^6$) ion in $Ca_3Co_2O_6$. In principle, the $(t_{2g})^5$ configuration can be approximated by either $(1a)^1(1e)^4$ or $(1a)^2(1e)^3$ (see **Fig. 2b**). The angular momentum behavior of the 1a (i.e., $z^2$) orbital is described by $d_0$, and those of the 1e orbitals by linear combinations of $d_{\pm 1}$ and $d_{\pm 2}$, namely, by $\sqrt{2/3}(x^2-y^2)-\sqrt{1/3}yz$ and $\sqrt{2/3}xy-\sqrt{1/3}xz$.[42] Thus, the orbital moment $\mu_L$ of the OCT $M^{4+}$ ($d^5$) ion would be negligible if its electron configuration is close to $(1a)^1(1e)^4$. However, this would not be the case if the electron configuration is close to $(1a)^2(1e)^3$. As discussed below, it depends on the spin arrangement between adjacent $Co^{2+}$ and $M^{4+}$ ions, the direct metal-metal interaction between them, and the SOC of the $M^{4+}$ ion whether the local electronic structure of the OCT $M^{4+}$ ($d^5$) ion is close to $(1a)^1(1e)^4$ or to $(1a)^2(1e)^3$.

As shown by the projected DOS plots for the FM state of $Ca_3CoRhO_6$ in **Fig. 10**, the LSDA+U+SOC (WIEN2k) calculations with $U_{eff}$(Co) = $U_{eff}$(Rh) = 4 eV predict $Ca_3CoRhO_6$ to be a magnetic insulator, whereas our LSDA+U calculations with $U_{eff}$(Co) = $U_{eff}$(Rh) = 4 eV predict $Ca_3CoRhO_6$ to be a metal (See **Fig. S1** of the supporting information). (The LSDA+ U and LSDA+U+SOC calculations with $U_{eff}$(Co) = 4 eV and $U_{eff}$(Rh) = 2 eV both predict $Ca_3CoRhO_6$ to be a metal. See **Fig. S2** of the supporting information). The projected DOS plots from the LSDA+U+SOC calculations show that the local electronic structure of the TP $Co^{2+}$ ion is given by $(z^2)^2(x^2-y^2, xy)^3(xz, yz)^2$, and that of the OCT $Rh^{4+}$ ion by $(1a)^1(1e)^4$. This explains the uniaxial magnetism of Ca$_3$CoRhO$_6$ brought about by the L = 2 configuration of the TP Co$^{2+}$ ion, and why the orbital moment μ$_L$ of the OCT Rh$^{4+}$ ion is nearly zero (see **Table 3**) in the LSDA+U+SOC calculations.

The above discussion for the FM state of Ca$_3$CoRhO$_6$, which accounts for the configuration $(z^2)^2(x^2-y^2, xy)^3(xz, yz)^2$ for the TP Co$^{2+}$ ion and the configuration $(1a)^1(1e)^4$ for the OCT Rh$^{4+}$ ion, is also applicable to the FM state of Ca$_3$CoIrO$_6$. The AFM spin arrangement (i.e., the ferrimagnetic state) of Ca$_3$CoIrO$_6$ has a slightly different picture for the local electronic structure of the OCT Ir$^{4+}$ ion. The projected DOS plots for the AFM state are presented in **Fig. 11**, where the OCT Ir$^{4+}$ ion is not described by $(1a)^1(1e)^4$ but by $(1a)^2(1e)^3$. (The LSDA+ U and LSDA+U+SOC calculations with U$_{eff}$(Co) = 4 eV and U$_{eff}$(Ir) = 2 eV both predict Ca$_3$CoIrO$_6$ to be a metal. See **Fig. S3** of the supporting information.) As a consequence, the orbital moment μ$_L$ of the OCT Ir$^{4+}$ ion is large (**Table 4**). This observation is explained by noting from **Fig. 9** that the σ*↓ level of the dimer made up of adjacent TP Co$^{2+}$ and OCT Ir$^{4+}$ ions would lie lower in energy in the AFM than in the FM spin arrangement, because the energy gap between the z$^2$↓ orbitals of the two ions is greater for the AFM arrangement. The 1e↓ level of the OCT Ir$^{4+}$ ion is split by SOC, and the OCT Ir$^{4+}$ ion adopts the $(1a)^2(1e)^3$ configuration when the upper one of the split 1e↓ level becomes higher in energy than the σ*↓ level (**Fig. 12**). Since the Ir$^{4+}$ ion has a stronger SOC than does the Rh$^{4+}$ ion, the split of the 1e↓ level is larger in Ca$_3$CoIrO$_6$ than in Ca$_3$CoRhO$_6$. In addition, the Co…Ir metal-metal interaction is weaker than the Co…Rh metal-metal interaction because the NN Co…Ir distance is longer than the NN Co…Rh distance. This makes the 1e↓ level lying lower in $Ca_3CoIrO_6$ than in $Ca_3CoRhO_6$. Consequently, the OCT $Rh^{4+}$ ion has the $(1a)^1(1e)^4$ configuration, but the OCT $Ir^{4+}$ ion the $(1a)^2(1e)^3$ configuration.

**7. Discussion**

For the SOC effect to induce electron transfer from the $\sigma^*\downarrow$ level to the $d_2\downarrow$ level in $Ca_3Co_2O_6$ (**Fig. 7a**), the $\sigma^*\downarrow$ level should lie high in energy because the split between the $d_2$ and $d_{-2}$ levels by SOC is not large for a 3d transition metal ion. Important factors raising the $\sigma^*\downarrow$ level are the direct metal-metal interaction and the FM spin arrangement between adjacent $Co^{3+}$ ions. Compared with the AFM arrangement (**Fig. 7b**), the FM arrangement has a smaller energy difference between the $z^2\downarrow$ orbitals of the TP and OCT $Co^{3+}$ ions, which leads to a stronger interaction between them hence raising the $\sigma^*\downarrow$ level higher. Another important factor is that the TP and OCT sites both have $Co^{3+}$ ions with similarly contracted $z^2\downarrow$ orbitals so that the overlap between them is good hence raising the $\sigma^*\downarrow$ level.

In the LSDA+U+SOC calculations, both the $\sigma\uparrow$ and $\sigma^*\uparrow$ levels are both occupied. In contrast, the $\sigma\downarrow$ level is filled but the $\sigma^*\downarrow$ level is not. Consequently, the Co…Co metal-metal interaction in $Ca_3Co_2O_6$ is overall bonding. This accounts for why the displacement of the TP $Co^{3+}$ ion is not large in the JT distorted structure of $Ca_3Co_2O_6$. This reasoning suggests that the Co…M metal-metal interaction in $Ca_3CoMO_6$ (M = Rh, Ir) should be weak because the TP $Co^{2+}$ ion has a large displacement in the JT distorted structure. The NN Co…Rh and Co…Ir distances of $Ca_3CoRhO_6$ and $Ca_3CoIrO_6$, respectively, are short (i.e., 2.682 and 2.706 Å, respectively) but are longer than the NN

Co…Co distance (2.595 Å) of $Ca_3Co_2O_6$. Furthermore, the Co 3d and Rh 4d orbitals are different in orbital contractedness, and even more so are the Co 3d and Ir 5d orbitals. Consequently, the direct metal-metal interaction between $Co^{2+}$ and $M^{4+}$ ions in $Ca_3CoMO_6$ (M = Rh, Ir) would be weaker than that between $Co^{3+}$ ions in $Ca_3Co_2O_6$. This accounts for why the displacement of the TP $Co^{2+}$ ion is large in the JT distorted structures of $Ca_3CoMO_6$ (M = Rh, Ir).

The differences in the $z^2$ orbital occupations of the TP and OCT ions in $Ca_3CoMO_6$ (M = Co, Rh, Ir) are important to note. From the viewpoint of two adjacent TP and OCT ions, the highest-lying level resulting from their two $z^2$ orbitals is the $\sigma^*\downarrow$ level, which decreases in energy with lengthening the NN Co…M distance and with increasing the difference in the contractedness of the Co and M $z^2$ orbitals. Thus, it is understandable that the two $z^2$ orbitals of adjacent TP and OCT ions have four electrons in $Ca_3CoIrO_6$ (i.e., the $\sigma^*\downarrow$ level is occupied), but three electrons in $Ca_3Co_2O_6$ and $Ca_3CoRhO_6$ (i.e., the $\sigma^*\downarrow$ level is unoccupied). The latter is equivalent to a singly occupied $z^2$ orbital at the TP $Co^{3+}$ ion in $Ca_3Co_2O_6$, but that at the OCT $Rh^{4+}$ ion in $Ca_3CoRhO_6$, due to the unequal weights of the TP and OCT $z^2$ orbitals in the $\sigma^*\downarrow$ level (**Fig. 7** vs. **Fig. 9**). A higher-lying $\sigma^*\downarrow$ level and a lower-lying $\sigma\downarrow$ level are obtained when adjacent TP and OCT ions have an FM spin arrangement than an AFM spin arrangement. Thus, the FM arrangement is energetically more favorable when the $\sigma^*\downarrow$ level is unoccupied as found for $Ca_3Co_2O_6$ and $Ca_3CoRhO_6$, but an AFM arrangement is energetically more favorable when the $\sigma^*\downarrow$ level is occupied as found for $Ca_3CoIrO_6$.

## 8. Concluding remarks

In summary, the JT instability, uniaxial magnetism, spin arrangement, metal-metal interaction and spin-orbit coupling are intimately interrelated in $Ca_3CoMO_6$ (M = Co, Rh, Ir). The adjacent spins in each $CoMO_6$ chain of $Ca_3CoMO_6$ (M = Co, Rh, Ir) prefer the FM arrangement for M = Co and Rh, but the AFM arrangement for M = Ir. The magnetism of $Ca_3CoMO_6$ (M = Co, Rh, Ir) cannot be genuinely uniaxial because it undergoes a weak JT distortion. Nevertheless, the orbital moments of the TP $Co^{n+}$ ions, though strongly reduced by the distortion, are still substantial enough to produce strong easy-axis anisotropy along the chain direction. The d-state split pattern of the TP $Co^{n+}$ (n = 2, 3) ions that is consistent with the electronic and magnetic properties of $Ca_3CoMO_6$ (M = Co, Rh, Ir) is not $(d_2, d_{-2}) < d_0 < (d_1, d_{-1})$, but $d_0 < (d_2, d_{-2}) < (d_1, d_{-1})$. The L = 2 configuration $(d_0)^1(d_2, d_{-2})^3(d_1, d_{-1})^2$ of the TP $Co^{3+}$ ion in $Ca_3Co_2O_6$ is a combined consequence of the FM spin arrangement between adjacent TP and OCT $Co^{3+}$ ions, the direct metal-metal interaction between them mediated by their $z^2$ orbials, and the SOC of the TP $Co^{3+}$ ion. In contrast to the case of $Ca_3Co_2O_6$, the TP and OCT ions of $Ca_3CoMO_6$ (M = Rh, Ir) have different oxidation states (+2 and +4, respectively), because the Co 3d orbital lies lower in energy, and is more contracted, than the Rh 4d and Ir 5d orbitals. The OCT $M^{4+}$ ion has the $(1a)^1(1e)^4$ configuration for M = Rh but the $(1a)^2(1e)^3$ configuration for M = Ir. This difference reflects a combined consequence of the spin arrangement between adjacent TP $Co^{2+}$ and OCT $M^{4+}$ ions, the direct metal-metal interaction between them mediated by their $z^2$ orbitals, and the SOC of the TP $M^{4+}$ ions.

**Acknowledgments**

The work at North Carolina State University was supported by the Office of Basic Energy Sciences, Division of Materials Sciences, U. S. Department of Energy, under Grant DE-FG02-86ER45259.

**References**

(1) Dai, D.; Whangbo, M.-H., *Inorg. Chem.* **2005**, *44*, 4407.

(2) (a) Kugel, K. I.; Khomskii, D. I., *Sov. Phys. Usp.* **1982**, *25*, 231. (b) Bersuker, I. B., *The Jahn-Teller Effect*; Cambridge University Press, 2006.

(3) (a) Reiff, W. M.; LaPointe, A. M.; Witten, E. H., *J. Am. Chem. Soc.* **2004**, *126*, 10206. (b) Reiff, W. M.; Schulz, C. E.; Whangbo, M.-H.; Seo, J. I.; Lee, Y. S.; Potratz, G. R.; Spicer, C. W.; Girolami, G. S., *J. Am. Chem. Soc.* **2009**, *131*, 404.

(4) Zhang, Y.; Xiang, H. J.; Whangbo, M.-H., *Phys. Rev. B* **2009**, *79*, 054432.

(5) Zubkov, V. G.; Bazuev, G. V.; Tyutyunnik, A. P.; Berger, I. F., *J. Solid State Chem.* **2001**, *160*, 293.

(6) Choi, Y. J.; Yi, H. T.; Lee, S.; Huang, Q.; Kiryukhin, V.; Cheong, S.-W., *Phys. Rev. Lett.* **2008**, *100*, 047601.

(7) Fjellvåg, H.; Gulbrandsen, E.; Åasland, S.; Olsen, A.; Hauback, B. C., *J. Solid State Chem.* **1996**, *124*, 190.

(8) Niitaka, S.; Kageyama, H.; Kato, M.; Yoshimura, K.; Kosuge, K., *J. Solid State Chem.* **1999**, *146*, 137.

(9) Kageyama, H.; Yoshimura, K.; Kosuge, K., *J. Solid State Chem.* **1998**, *140*, 14.

(10) (a) Darriet, J.; Subramanian, M. A., *J. Mater. Chem.* **1995**, *5*, 543. (b) Perez-Mato, J. M.; Zakhour-Nakhl, M.; Weill, F.; Darriet, J. *J. Mater. Chem.* **1999**, *9*, 2795. (c)

Stitzer, K. E.; Darriet. J.; Zur Loye, H. -C. *Curr. Opin. Solid State. Mater. Sci.* **2001**, *5*, 535.

(11) Takubo, K.; Mizokawa, T.; Hirata, S.; Son, J.-Y.; Fujimori, A.; Topwal, D.; Sarma, D. D.; Rayaprol, S.; Sampathkumaran, E.-V., *Phys. Rev. B* **2005**, *71*, 073406.

(12) Åasland, S.; Fjellvåg, H.; Hauback, B., *Solid State Commun*. **1997**, *101*, 187.

(13) Kageyama, H.; Yoshimura, K.; Kosuge, K.; Azuma, M.; Takano, M.; Mitamura, H.; Goto, T., *J. Phys. Soc. Jpn*. **1997**, *66*, 3996.

(14) Maignan, A.; Michel, C.; Masset, A. C.; Martin, C.; Raveau, B., *Eur. Phys. J. B* **2000**, *15*, 657.

(15) (a) Niitaka, S.; Kageyama, H.; Yoshimura, K.; Kosuge, K.; Kawano, S.; Aso, N.; Mitsuda, A.; Mitamura, H.; Goto, T., *J. Phys. Soc. Jpn*. **2001**, *70*, 1222. (b) Niitaka, S.; Yoshimura, K.; Kosuge, K.; Nishi, M.; Kakurai, K., *Phys. Rev. Lett*. **2001**, *87*, 177202.

(16) Sugiyama, J.; Morris, G. D.; Nozaki, H.; Ikedo, Y.; Russo, P. L.; Stubbs, S. L.; Brewer, J. H.; Ansaldo, E. J.; Martin, C.; Hébert, S.; Maignan, A., *Physica B* **2009**, *404*, 603.

(17) Whangbo, M.-H.; Dai, D.; Koo, H.-J.; Jobic, S., *Solid State Commun*. **2003**, *125*, 413.

(18) Vidya, R.; Ravindran, P.; Fjellvåg, H.; Kjekshus, A., *Phys. Rev. Lett*. **2003**, *91*, 186404.

(19) Eyert, V.; Laschinger, C.; Kopp, T.; Frésard, R., *Chem. Phys. Lett*. **2004**, *385*, 249.

(20) Vidya, R.; Ravindran, P.; Vajeeston, P.; Fjellvåg, H.; Kjekshus, A., *Ceramics Intern*. **2004**, *30*, 1993.

(21) Wu, H.; Haverkort, M. W.; Hu, Z.; Khomskii, D. I.; Tjeng, L. H., *Phys. Rev. Lett.* **2005**, *95*, 186401.

(22) Villesuzanne, A.; Whangbo, M.-H., *Inorg. Chem.* **2005**, *44*, 6339.

(23) Stoeffler, D., *Microelectronic Engineering*, **2008**, *85*, 2451.

(24) Eyert, V.; Schwingenschlögl, U.; Hackenberger, C.; Kopp, T.; Frésard, R.; Eckern, U., *J. Solid State Chem.* **2007**, *36*, 156.

(25) Wu, H.; Hu, Z.; Khomskii, D. I.; Tjeng, L. H., *Phys. Rev. B* **2007**, *75*, 245118.

(26) (a) Stiefel, E. I.; Eisenberg, R.; Rosenberg, R. C.; Gray, H. B., *J. Am. Chem. Soc.* **1966**, *88*, 2956. (b) Schrauzer, G. N.; Mayweg, V. P., *J. Am. Chem. Soc.* **1966**, *68*, 3234. (c) Hulliger, F., *Struct. Bonding (Berlin)*, **1968**, *4*, 83. (d) Anzenhofer, K.; van den Berg, J. M.; Cossee, P.; Heile, J. N., *J. Phys. Chem. Solids* **1970**, *31*, 1057. (e) Hoffmann, R.; Howell, J. M.; Rossi, A. R., *J. Am. Chem. Soc.* **1976**, *98*, 2484.

(27) Burnus, T.; Hu, Z.; Haverkort, M. W.; Cezar, J. C.; Flahaut, D.; Hardy, V.; Maignan, A.; Brookes, N. B.; Tanaka, A.; Hsieh, H. H.; Lin, H.-J.; Chen, C. T.; Tjeng, L. H., *Phys. Rev. B* **2006**, *74*, 245111.

(28) Albright, T. A, Burdett, J. K., Whangbo, M.-H., *Orbital Interactions in Chemistry*: Wiley; New York, 1985.

(29) Burnus, T.; Hu, Z.; Wu, H.; Cezar, J. C.; Niitaka, S.; Takagi, H.; Chang, C. F.; Brookes, N. B.; Lin, H.-J.; Jang, L. Y.; Tanaka, A.; Liang, K. S.; Chen, C. T.; Tjeng, L. H., *Phys. Rev. B* **2008**, *77*, 205111.

(30) (a) Kresse, G.; Hafner, J., *Phys. Rev. B* **1993**, *47*, 558. (b) Kresse, G.; Furthmüller, J., *Comput. Mater. Sci.* **1996**, *6*, 15. (c) Kresse, G.; Furthmüller, J., *Phys. Rev. B* **1996**, *54*, 11169.

(31) Dudarev, S. L.; Botton, G. A.; Savrasov, S. Y.; Humphreys, C. J.; Sutton, A. P., *Phys. Rev. B* **1998**, *57*, 1505.

(32) Kuneš, J.; Novák, P.; Diviš, M.; Oppeneer, P. M., *Phys. Rev. B* **2001**, *63*, 205111.

(33) Kageyama, H.; Yoshimura, K.; Kosuge, K., *J. Solid State Chem.* **1998**, *140*, 14.

(34) Singh, D. J. *Plane waves, Pseudopotentials and the LAPW Method*: Kluwer Academic; Boston, 1994.

(35) Blaha, P.; Schwarz, K.; Madsen, G. K. H.; Kvasnicka, D.; Luitz, J. *WIEN2K, An Augmented Plane Wave + Local Orbitals Program for Calculating Crystal Properties* (Techn. Universität Wien, Austria, 2001).

(36) Whangbo, M.-H.; Koo, H.-J.; Villesuzanne, A.; Pouchard, M., *Inorg. Chem.* **2002**, *41*, 1920.

(37) Hoffmann, R., *J. Chem. Phys.* **1963,** *39,* 1397.

(38) Our calculations were carried out by employing the SAMOA (Structure and Molecular Orbital Analyzer) program package (This program can be downloaded free of charge from the website, http://chvamw.chem.ncsu.edu/).

(39) Mapps, F. E.; Machin, D. J., *Magnetism and Transition Metal Complexes*: Chapman and Hall, London, 1973.

(40) Dai, D.; Xiang, H. J.; Whangbo, M.-H., *J. Comput. Chem.* **2008**, *29*, 2187.

(41) Clementi, E.; Roetti, C. *Atomic Data Nuclear Data Tables* **1974**, *14*, 177. (b)

(42) (a) Albright, T. A.; Hofmann, P.; Hoffmann, R., *J. Am. Chem. Soc.* **1977**, *99*, 7546. (b) Orgel, L. E., *An Introduction to Transition Metal Chemistry*: Wiley, New York, 1969, p 174.

Table 1.    Relative energies ΔE (meV/FU) of the experimental and optimized structures of $Ca_3CoMO_6$ (M = Co, Rh, Ir) obtained from the LSDA+U+SOC calculations using the PAW method of the VASP with $U_{eff}$ = 4 eV for Co and $U_{eff}$ = 2 eV for M = Rh and Ir.[a]

| Geometry used | $Ca_3Co_2O_6$ | $Ca_3CoRhO_6$ | $Ca_3CoIrO_6$ |
|---|---|---|---|
| Experimental with $C_3$ axis | 94.3 | 65.7 | - |
| Optimized with $C_3$ axis | 65.5 | 35.2 | 139.8 |
| Optimized with no $C_3$ axis | 0.0 | 0.0 | 0.0 |

[a] For each $Ca_3CoMO_6$ (M = Co, Rh, Ir), the optimization was carried out for the FM state.

Table 2. Spin and orbital moments ($\mu_S$ and $\mu_L$, respectively) of the TP and OCT $Co^{3+}$ ions in the FM state of $Ca_3Co_2O_6$ obtained from the LSDA+U+SOC (WIEN2k)[a] calculations with $U_{eff}(Co) = 4$ eV.[b,c]

| Geometry | $Co^{3+}$ (TP) | | $Co^{3+}$ (OCT) | |
|---|---|---|---|---|
| | $\mu_S$ ($\mu_B$) | $\mu_L$ ($\mu_B$) | $\mu_S$ ($\mu_B$) | $\mu_L$ ($\mu_B$) |
| Experimental with $C_3$ axis | 2.94 | 1.58 | 0.08 | 0.18 |
| Optimized with $C_3$ axis[a] | 2.94/2.94 | 1.48/1.48 | 0.16/0.16 | 0.05/0.05 |
| Optimized with no $C_3$ axis | 2.92/2.89 | 0.31/0.45 | 0.02/0.02 | 0.03/0.02 |

[a] Our LSDA+U+SOC (WIEN2k) optimization converges to the structure with no $C_3$-rotational symmetry. The numbers listed are obtained from our LSDA+U+SOC (VASP) optimization.

[b] The orbital and spin moments have the same direction when they have the same sign, and the opposite directions otherwise.

[c] There are two slightly different TP Co atoms as well as two slightly different OCT Co atoms in the optimized structures with or without $C_3$ symmetry.

Table 3. Spin and orbital moments ($\mu_S$ and $\mu_L$, respectively) of the TP $Co^{2+}$ and OCT $Rh^{4+}$ ions in the FM state of $Ca_3CoRhO_6$ obtained from the LSDA+U+SOC (WIEN2k) calculations with $U_{eff}$ (Co) = $U_{eff}$ (Rh) = 4 eV.[a,b]

| Geometry | $Co^{2+}$ (TP) | | $Rh^{4+}$ (OCT) | |
|---|---|---|---|---|
| | $\mu_S$ ($\mu_B$) | $\mu_L$ ($\mu_B$) | $\mu_S$ ($\mu_B$) | $\mu_L$ ($\mu_B$) |
| Experimental with $C_3$ axis | 2.71 | 1.76 | 0.49 | 0.01 |
| Optimized with $C_3$ axis | 2.69/2.69 | 1.76/1.76 | 0.59/0.59 | 0.01/0.01 |
| Optimized with no $C_3$ axis | 2.64/2.64 | 0.50/0.50 | 0.31/0.31 | 0.01/0.01 |

[a] The orbital and spin moments have the same direction when they have the same sign, and the opposite directions otherwise.

[b] There are two slightly different TP Co atoms as well as two slightly different OCT Co atoms in the optimized structures with or without $C_3$ symmetry.

Table 4.  Spin and orbital moments ($\mu_S$ and $\mu_L$, respectively) of the TP $Co^{2+}$ and OCT $Ir^{4+}$ ions in the AFM state of $Ca_3CoIrO_6$ obtained from the LSDA+U+SOC (WIEN2k) calculations with $U_{eff}$ (Co) = $U_{eff}$ (Ir) = 4 eV.[a,b]

| Geometry used | $Co^{2+}$ (TP) | | $Ir^{4+}$ (OCT) | |
|---|---|---|---|---|
| | $\mu_S$ ($\mu_B$) | $\mu_L$ ($\mu_B$) | $\mu_S$ ($\mu_B$) | $\mu_L$ ($\mu_B$) |
| Optimized with $C_3$ axis | 2.62/2.62 | 1.77/1.77 | −0.43/−0.43 | −0.51/−0.51 |
| Optimized with no $C_3$ axis | 2.62/2.62 | 0.57/0.72 | −0.44/−0.44 | −0.54/−0.54 |

[a] The orbital and spin moments have the same direction when they have the same sign, and the opposite directions otherwise.

[b] There are two slightly different TP Co atoms as well as two slightly different OCT Co atoms in the optimized structures with or without $C_3$ symmetry.

**Figure captions**

Figure 1.   (a) Projection view of the crystal structure of $Ca_3CoMO_6$ along the c-direction. (b) Perspective view of an isolated $CoMO_6$ chain. The grey, purple, blue and red balls represent Ca, Co, M, and O atoms, respectively.

Figure 2.   Shapes and relative energies of (a) the d-states of the $CoO_6$ trigonal prism and (b) the $t_{2g}$-states of the $CoO_6$ octahedron of $Ca_3Co_2O_6$ obtained from extended Hückel tight-binding calculations.

Figure 3.   High-spin electron configurations expected for the $Co^{3+}$ ($d^6$) and $Co^{2+}$ ($d^7$) ions at a trigonal prism site when the d-state split pattern is given by $d_0 < (d_2, d_{-2}) < (d_1, d_{-1})$ in (a), (b) and (e), and by $(d_2, d_{-2}) < d_0 < (d_1, d_{-1})$ in (c) and (d).

Figure 4.   Displacements of the atoms associated with the Jahn-Teller distortions in the magnetic ground state of (a) $Ca_3Co_2O_6$, (b) $Ca_3CoRhO_6$ and (c) $Ca_3CoIrO_6$ with respect to their positions of the experimental structures for $Ca_3Co_2O_6$ and $Ca_3CoRhO_6$, and with respect to their positions of the optimized structure with $C_3$ symmetry for $Ca_3CoIrO_6$. The largest atom displacement is 0.044 Å in $Ca_3Co_2O_6$, 0.064 Å in $Ca_3CoRhO_6$, and 0.051 Å in $Ca_3CoIrO_6$. In each figure, the left side shows a perspective view of the atom displacements in the $CoMO_6$ chain, and the right side the projection view (along the chain direction) of the atom displacements in the $CoO_6$ trigonal prisms and $MO_6$ octahedra.

29Figure 5.   Projected DOS plots for the $z^2$, $(x^2-y^2 + xy)$ and $(xz + yz)$ states of the TP and OCT $Co^{3+}$ ions in the FM state of $Ca_3Co_2O_6$ obtained from the LSDA+U and LSDA+U+SOC calculations by using the FPLAPW method of the WIEN2k package, the experimental structure of $Ca_3Co_2O_6$, and $U_{eff}(Co) = 4$ eV.

Figure 6.   Schematic representations of the high-spin electron configuration of the trigonal prism $Co^{3+}$ ($d^6$) in (a) the one-electron picture and (b) the spin-polarized DFT+U level of description.

Figure 7.   DFT+U level description of the orbital interactions between the $z^2$ orbitals of adjacent TP and OCT $Co^{3+}$ ions in $Ca_3Co_2O_6$ that lead to the σ and σ* orbitals when the spins of the two $Co^{3+}$ sites have a (a) ferromagnetic and (b) an antiferromagnetic arrangement.

Figure 8.   Effect of the SOC at the TP $Co^{3+}$ ion on the occupancy of the σ*↓ level of a dimer unit consisting of two adjacent TP and OCT $Co^{3+}$ ions in $Ca_3Co_2O_6$ for cases when the d-state split pattern of the TP $Co^{3+}$ ion is given by (a) $d_0 < (d_2, d_{-2}) < (d_1, d_{-1})$ and (b) $(d_2, d_{-2}) < d_0 < (d_1, d_{-1})$.

Figure 9.   Orbital interactions between the $z^2$↓ orbitals of adjacent TP $Co^{2+}$ and OCT $M^{4+}$ ions in $Ca_3CoMO_6$ (M = Rh, Ir) that lead to the σ↓ and σ*↓ orbitals when the spins of the two ion sites have (a) an FM and (b) an AFM arrangement. The midpoint between the $z^2\uparrow$ and $z^2\downarrow$ orbitals is higher in energy for the $M^{4+}$ (M = Rh, Ir) ion than that for the $Co^{2+}$ ion, because the Rh 4d and Ir 5d orbital is more diffuse, and lies higher in energy, than the Co 3d orbital. The $\sigma^*\downarrow$ orbital lies higher in energy in the FM than in the AFM spin arrangement.

Figure 10.  Projected DOS plots for the $z^2$, $(x^2-y^2 + xy)$ and $(xz + yz)$ states of the TP $Co^{2+}$ and OCT $Rh^{4+}$ ions in the FM state of $Ca_3CoRhO_6$ obtained from the LSDA+U+SOC calculations by using the FPLAPW method of the WIEN2k package, the experimental structure of $Ca_3CoRhO_6$, and $U_{eff}$ = 4 eV on both Co and Rh.

Figure 11.  Projected DOS plots for the $z^2$, $(x^2-y^2 + xy)$ and $(xz + yz)$ states of the TP $Co^{2+}$ and OCT $Ir^{4+}$ ions in the AFM state of $Ca_3CoIrO_6$ obtained from the LSDA+U+SOC calculations by using the FPLAPW method of the WIEN2k package, the experimental structure of $Ca_3CoIrO_6$, and $U_{eff}(Co) = U_{eff}(Ir) = 4$ eV.

Figure 12.  SOC effects on the $1e\downarrow$ level of the OCT $M^{4+}$ ion and on the occupancy of the $\sigma^*\downarrow$ level of a dimer made up of two adjacent TP $Co^{2+}$ and OCT $M^{4+}$ ions in $Ca_3CoMO_6$ (M = Rh, Ir): (a) $Ca_3CoRhO_6$ and (b) $Ca_3CoIrO_6$.

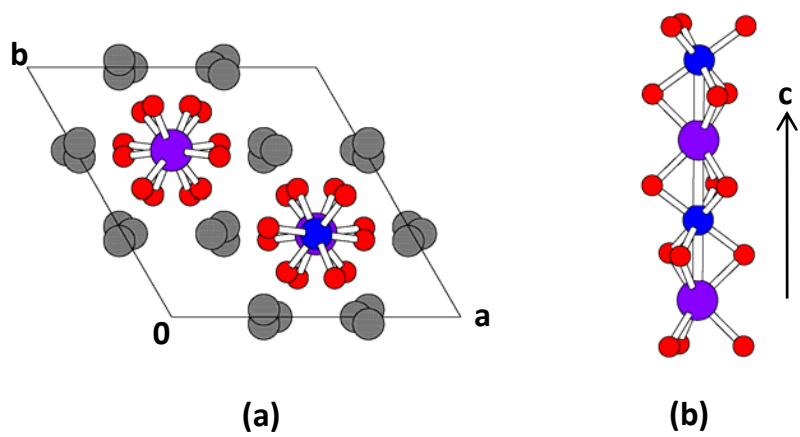

Figure 1.

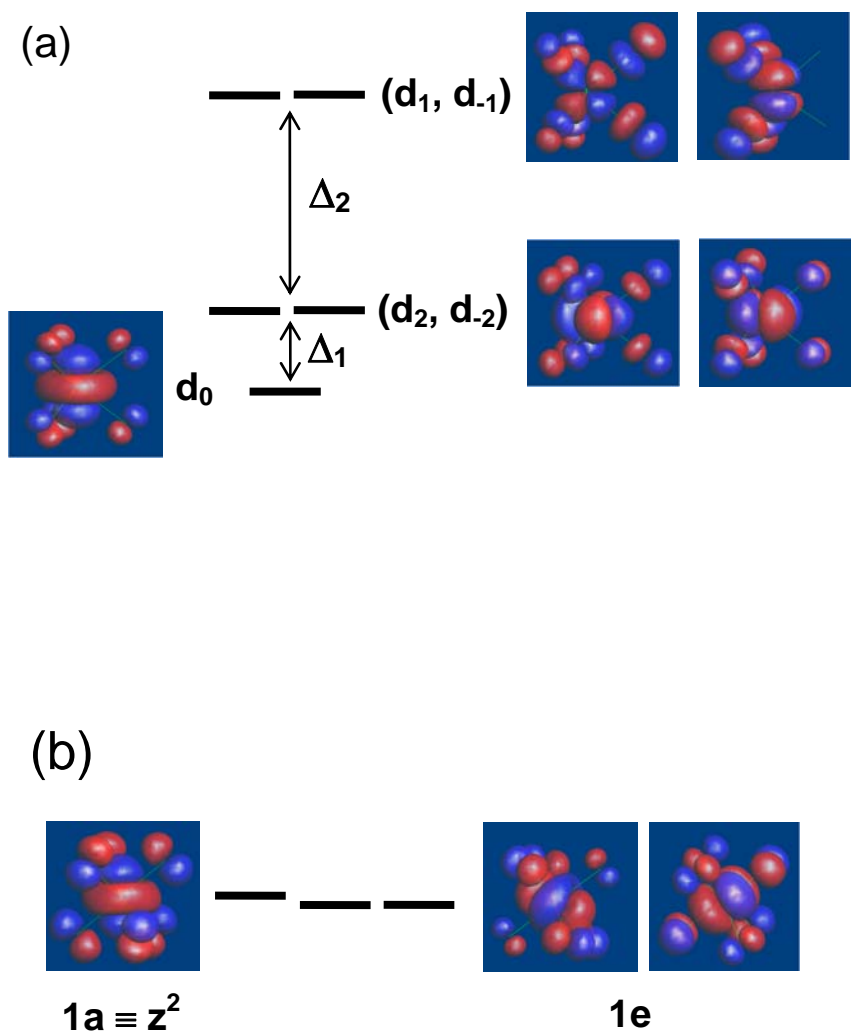

Figure 2

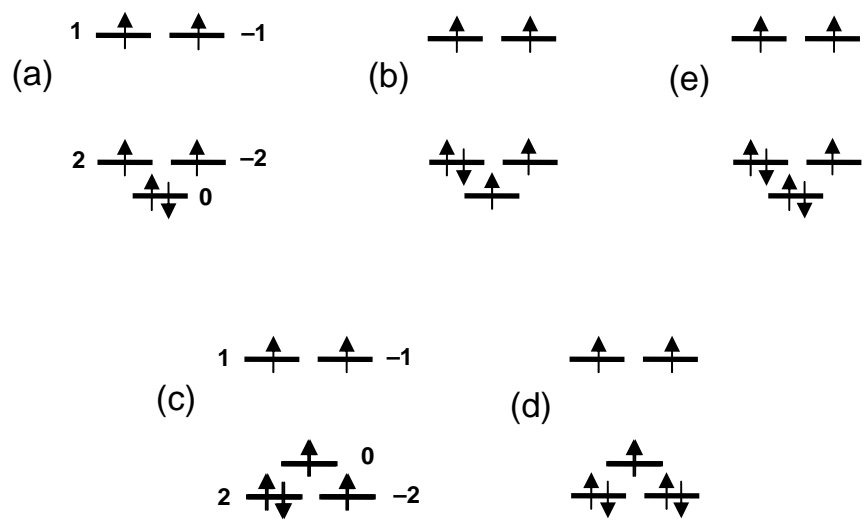

Figure 3

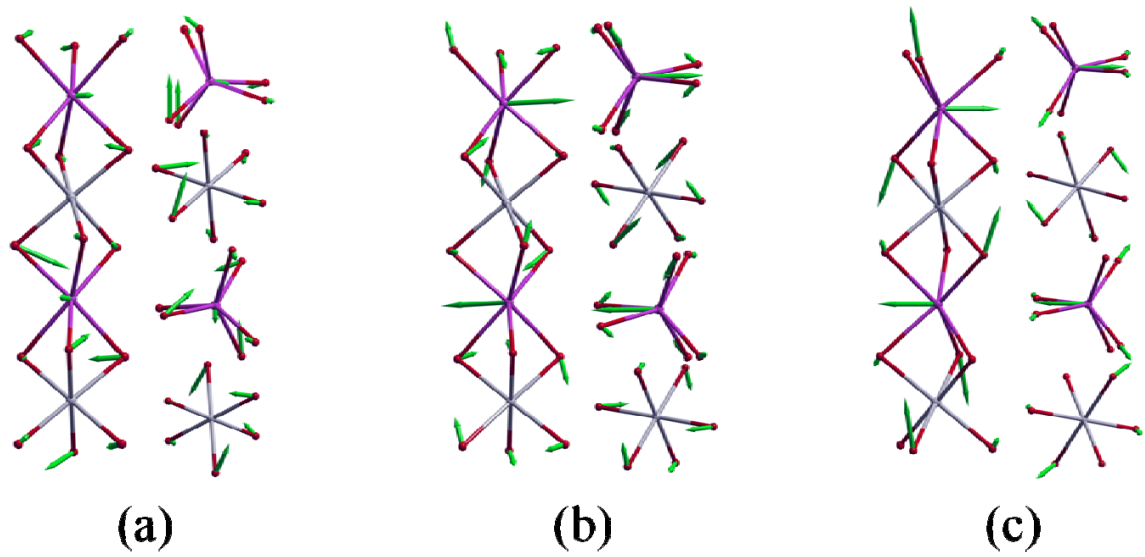

Figure 4

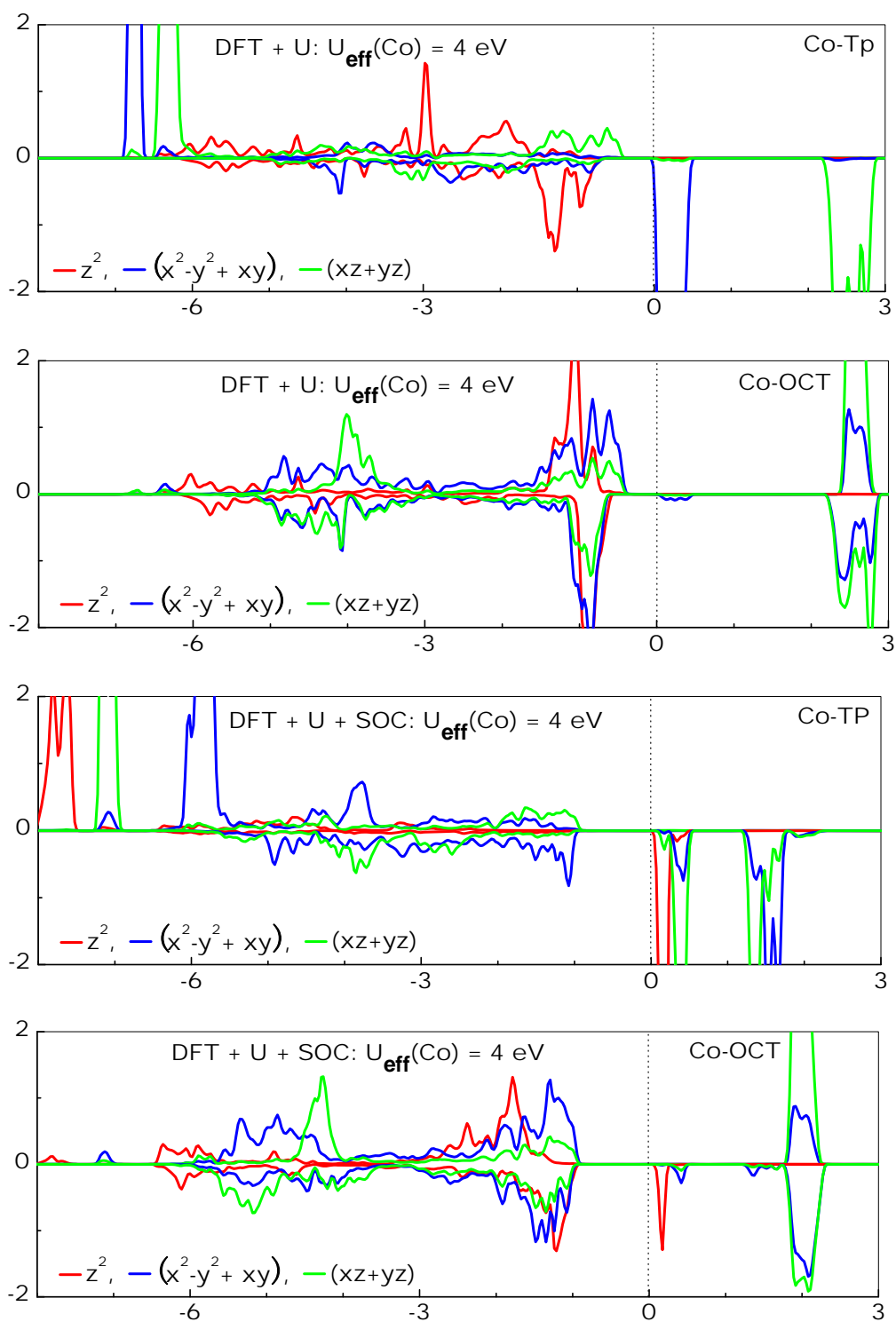

Figure 5

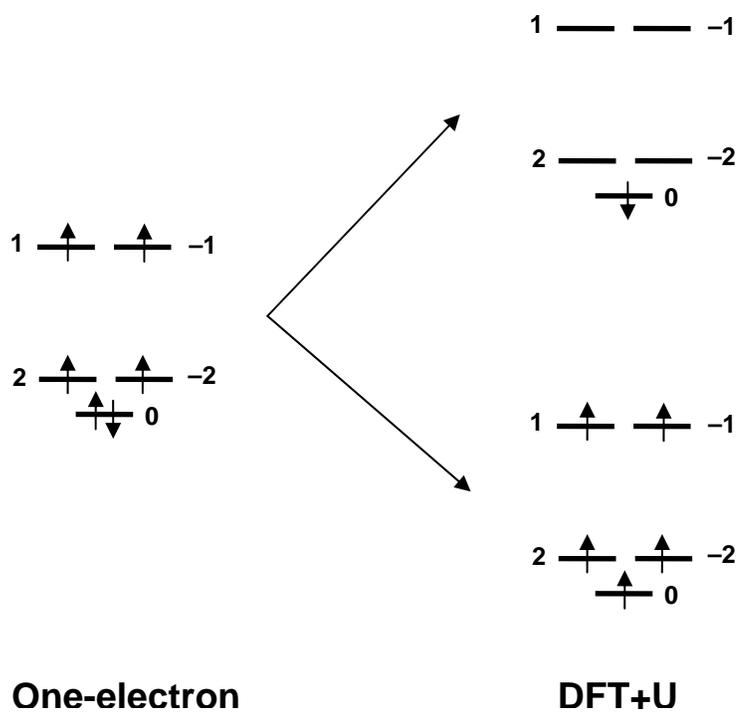

**One-electron**  **DFT+U**

Figure 6

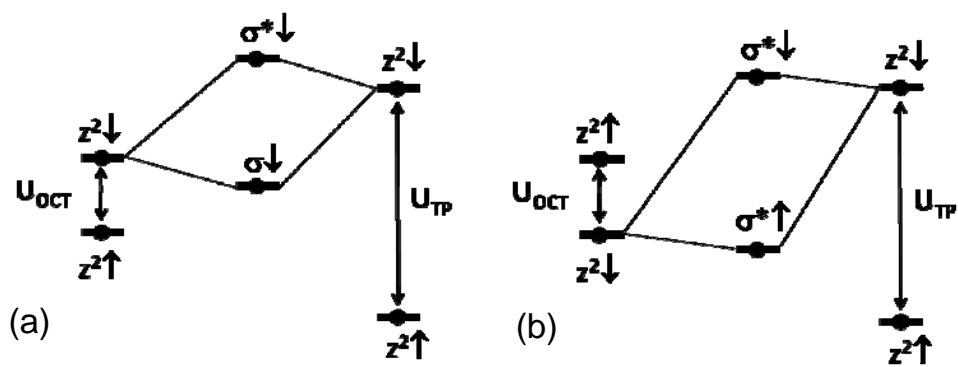

Figure 7

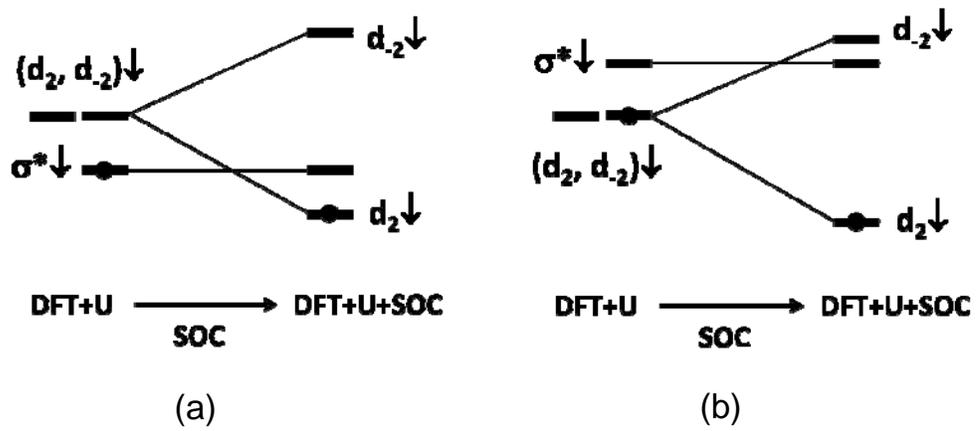

Figure 8

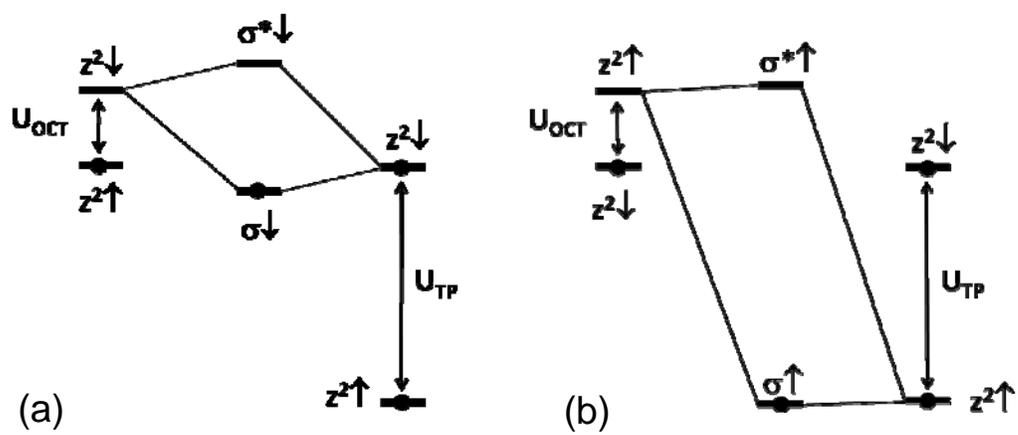

Figure 9

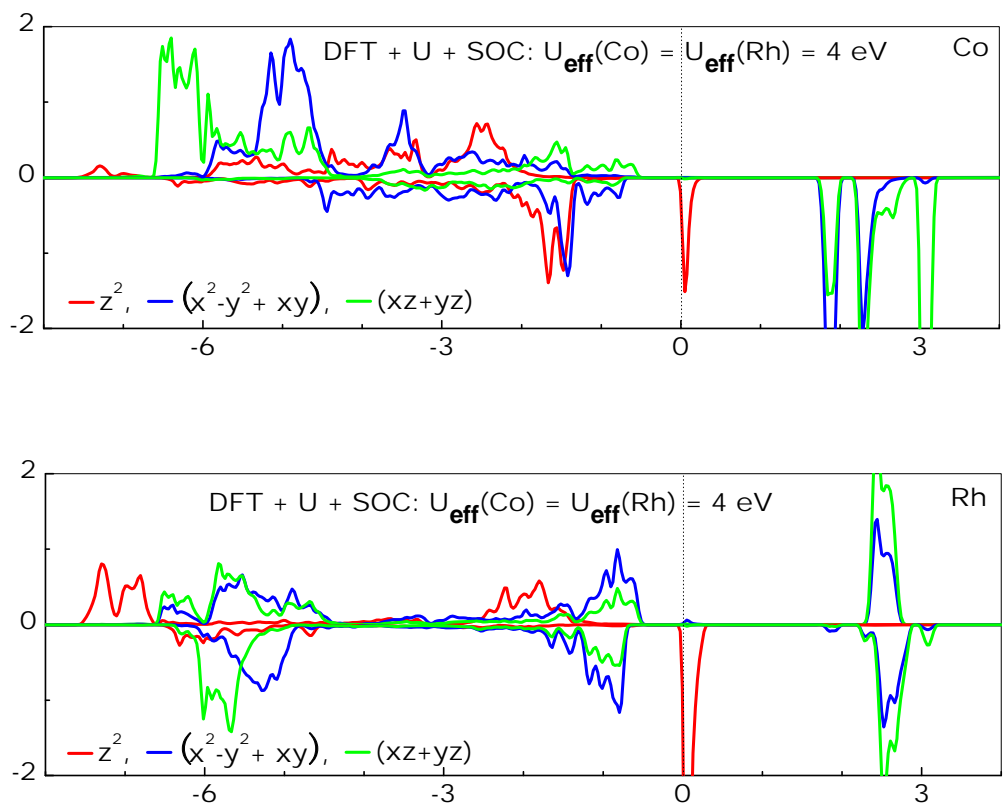

Figure 10

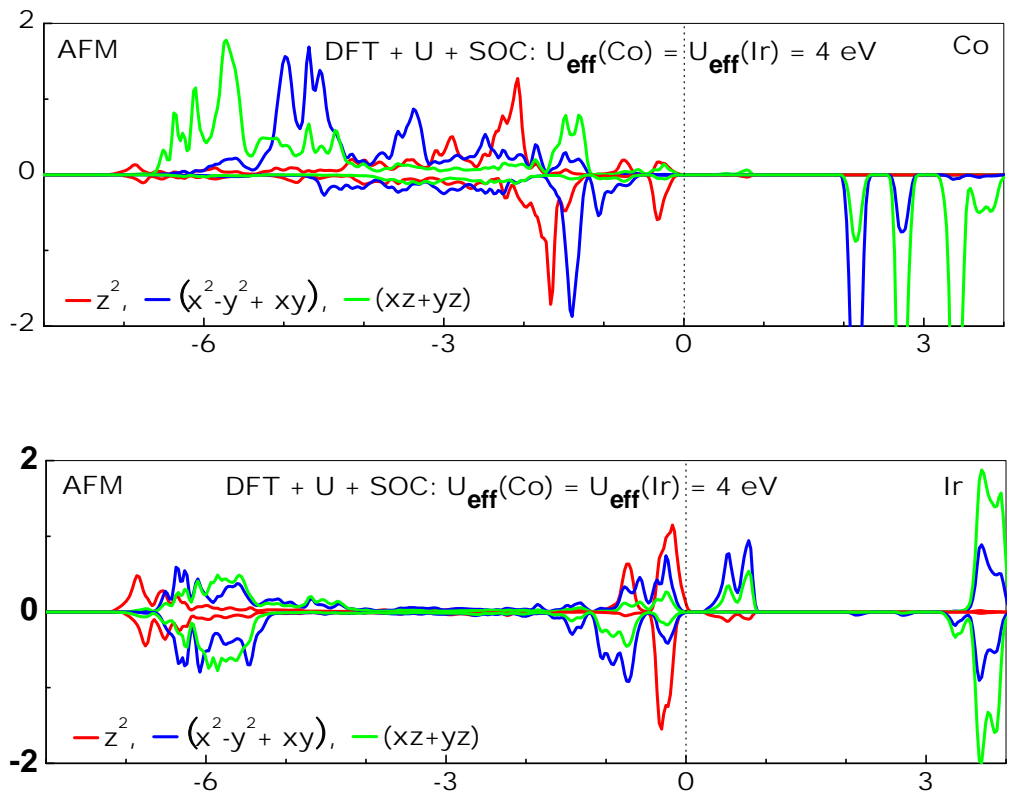

Figure 11

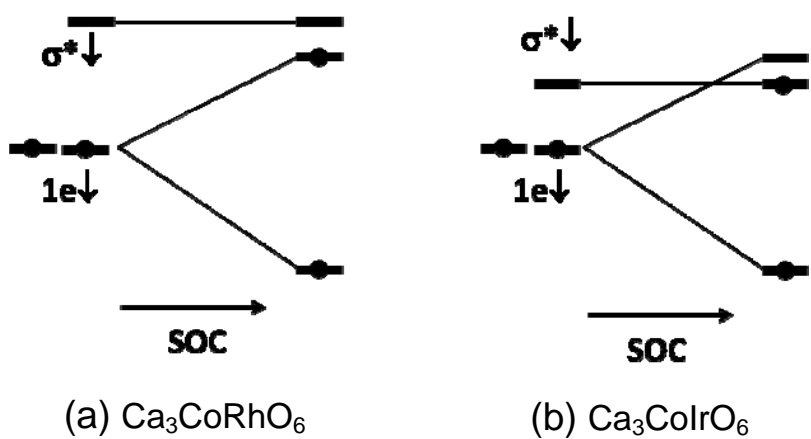

(a) $Ca_3CoRhO_6$  (b) $Ca_3CoIrO_6$

Figure 12.

**Synopsis**

We compared the magnetic and electronic properties of the three magnetic oxides $Ca_3CoMO_6$ (M = Co, Rh, Ir) on the basis of density functional theory calculations including on-site repulsion and spin-orbit coupling, and probed the essential one-electron pictures hidden behind results of these calculations. Our analysis reveals that the magnetic and electronic properties of these oxides are governed by an intimate interplay between Jahn-Teller instability, uniaxial magnetism, spin arrangement, direct metal-metal bonding, and spin-orbit coupling.

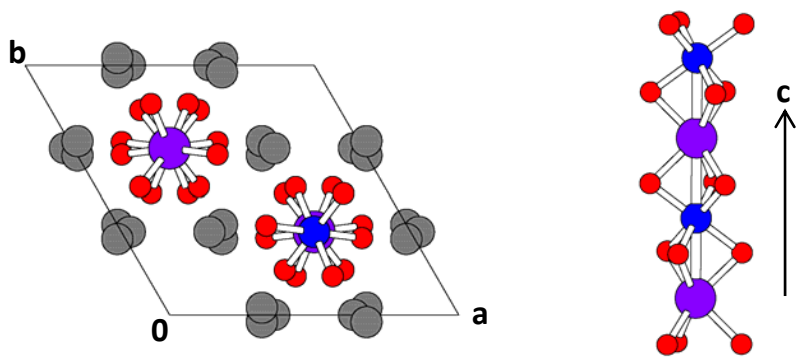